\documentclass[manuscript,a4paper,usenames,dvipsnames]{aastex61}

\usepackage{epstopdf}
\usepackage{amsmath} \usepackage{amsfonts} \usepackage{amssymb}
\usepackage{wasysym}
\usepackage{xcolor}

\renewcommand{\d}[1]{\mathinner{{\rm d}#1}}
\newcommand{\fn}[2]{\mathinner{#1\mathopen{\left(#2\right)}}}
\newcommand{\eq}[1]{Eq.~(\ref{#1})}
\newcommand{\eqnp}[1]{Eq.~\ref{#1}}

\newcommand{\pk}{\mathinner{P_{\rm m}(k)}}
\newcommand{\msun}{\mathinner{{\rm M}_{\odot}}}
\newcommand{\msunh}{\mathinner{h^{-1}{\rm M}_{\odot}}}

\newcommand{\Mpch}{\mathinner{h^{-1}{\rm Mpc}}}

\newcommand{\keV}{\mathinner{\mathrm{keV}}}

\newcommand{\GeV}{\mathinner{\mathrm{GeV}}}
\newcommand{\TeV}{\mathinner{\mathrm{TeV}}}
\newcommand{\invMpc}{\mathinner{{\rm Mpc}^{-1}}}
\newcommand{\invMpch}{\mathinner{h{\rm Mpc}^{-1}}}
\newcommand{\kmsMpc}{\mathinner{\mathrm{km/s/Mpc}}}

\newcommand{\kb}{\mathinner{k_{\rm b}}}
\newcommand{\Msupp}[1]{\mathinner{M_{\rm supp}^{\rm #1}}}
\newcommand{\khalf}{\mathinner{k_{1/2}}}
\newcommand{\mFD}{\mathinner{m_{\rm FD}}}

\newcommand{\dTb}{\mathinner{\delta T_{\rm b}}}

\newcommand{\fref}[1]{Figure~\ref{#1}}
\newcommand{\sref}[1]{Section~\ref{#1}}

\begin{document}

\title{Small-scale Effects of Thermal Inflation on Halo Abundance at High-$z$, Galaxy Substructure Abundance and 21-cm Power Spectrum}

\author{Sungwook E.~Hong}
\affiliation{Korea Astronomy and Space Science Institute, Daejeon 34055, Korea; swhong@kasi.re.kr}
\author{Heeseung Zoe}
\affiliation{School of Basic Science, Daegu Gyungbuk Institute of Science and Technology (DGIST), Daegu 42988, Korea; heezoe@dgist.ac.kr}
\author{Kyungjin Ahn}
\affiliation{Department of Earth Sciences, Chosun University, Gwangju 61452, Korea; Corresponding author; kjahn@chosun.ac.kr}

\date{\today}

\begin{abstract}
We study the impact of thermal inflation on the formation of cosmological structures and present astrophysical observables which can be used to constrain and possibly probe the thermal inflation scenario.
These are dark matter halo abundance at high redshifts, satellite galaxy abundance in the Milky Way, and fluctuation in the 21-cm radiation background before the epoch of reionization.
The thermal inflation scenario leaves a characteristic signature on the matter power spectrum by boosting the amplitude at a specific wavenumber determined by the number of e-foldings during thermal inflation ($N_{\rm bc}$), and strongly suppressing the amplitude for modes at smaller scales. For a reasonable range of parameter space, one of the consequences is the suppression of minihalo formation at high redshifts and that of satellite galaxies in the Milky Way. While this effect is substantial, it is degenerate with other cosmological or astrophysical effects. The power spectrum of the 21-cm background probes this impact more directly, and its observation may be the best way to constrain the thermal inflation scenario due to the characteristic signature in the power spectrum. The Square Kilometre Array (SKA) in phase 1 (SKA1) has sensitivity large enough to achieve this goal for models with $N_{\rm bc}\gtrsim 26$ if a 10000-hr observation is performed. The final phase SKA, with anticipated sensitivity about an order of magnitude higher, seems more promising and will cover a wider parameter space.
\end{abstract}

\section{Introduction}

Thermal inflation was introduced to primarily solve the moduli problem that is generic in supersymmetric cosmology \citep{Lyth:1995hj, Lyth:1995ka}.
Moduli particles can be long-lived and disturb Big Bang nucleosynthesis (BBN) as they decay \citep{Coughlan:1983ci, Banks:1993en, deCarlos:1993wie}.
Thermal inflation dilutes these unwanted particles by a low energy inflationary epoch, which is driven by thermal effects holding an unstable flat direction at the origin in supersymmetric theories  \citep{Lyth:1995hj,Lyth:1995ka,Yamamoto:1985mb,Yamamoto:1985rd,Enqvist:1985kz,Bertolami:1987xb,Ellis:1986nn,Ellis:1989ii,Randall:1994fr}.
Thermal inflation has a rich phenomenology in particle cosmology: 
it provides a mechanism for baryogenesis \citep{Stewart:1996ai,Jeong:2004hy,Felder:2007iz,Kim:2008yu,Kawasaki:2006py,Lazarides:1985ja,Yamamoto:1986jw,Mohapatra:1986dg} and has implications for axion/axino dark matter \citep{Yamamoto:1985mb,Yamamoto:1986jw,Kim:2008yu}.

Two kinds of cosmological or astrophysical venues for proving or constraining the thermal inflation scenario seem possible.
First, the matter power spectrum, which is an indicator of the clustering properties of the dark matter and baryons, can be strongly affected by thermal inflation.
As shown in an analytic calculation by \citet{Hong:2015oqa}, the curvature perturbation on constant energy density hypersurface at largest scales ($k\le k_{\rm b}$, where $k_{\rm b}$ is a model-dependent parameter defined in \sref{sec:theory}) remains outside the horizon until or after the matter domination era. 
Therefore, they are still compatible with the standard inflation models and can be well fit by the large-scale observations made by the {\em Wilkinson Microwave Anisotropy Probe} \citep[WMAP;][]{Planck:2013}, the {\em Planck} \citep{Spergel:2006}, and the {\em Sloan Digital Sky Survey}.
However, the perturbation at smaller scales with $k > k_{\rm b}$ is strongly suppressed, and the corresponding power spectrum is reduced by $\fn{\mathcal{O}}{10^2}$.
It therefore suggests that thermal inflation can be tested by small-scale observations regarding ultracompact minihalos or primordial black holes \citep{Carr:1975qj,Josan:2009qn,Bringmann:2011ut}, the lensing dispersion of SNIa \citep{Ben-Dayan:2013eza}, Lyman-$\alpha$ forest tomography \citep{Baur:2016}, CMB distortions \citep{Cho:2017,Chluba:2011hw,Chluba:2012gq} and the 21-cm hydrogen line background at or prior to the era of reionization \citep{Cooray:2006km, Mao:2008ug}. 
Second, thermal inflation produces a unique profile of gravitational wave background.
Thermal inflation suppresses the primordial gravitational wave at the scale of the solar system or smaller \citep{Mendes:1998gr}, but the first order phase transition at the end of thermal inflation changes the stochastic gravitational wave background by generating a new type of gravitational waves with frequencies in the Hz range \citep{Kosowsky:1991ua, Easther:2008sx}.
Though their amplitude is small---the peak of gravitational wave power is  $\fn{\Omega_\mathrm{GW}}{f \approx 1 \mathrm{Hz} }h^2 \sim 10^{-17}$---they are potentially detectable by {\em Big Bang Observer} \citep[BBO;][]{Crowder:2005nr}. 

In this paper, we only treat the first venue for the constraints, namely the impact of thermal inflation on the matter power spectrum and its detectability. 
Among various observables associated with this, we will provide quantitative predictions on the hydrogen 21-cm line background that can be probed by high-sensitivity radio telescopes such as the {\em Square Kilometre Array} \citep[SKA;][]{Furlanetto:2009qk}, the global halo abundance, and the abundance of satellite galaxies inside the Milky Way. 
One of the main science goals of this 21-cm background observations is to probe the power spectrum at redshifts much higher ($z \lesssim 30-10$ depending on radio telescopes) than those targeted by typical galaxy surveys ($z\lesssim2$). 
Possibility to probe the power spectrum at such high redshifts is favorable because some of the small-scale modes which became nonlinear at low redshifts may remain still in the linear regime at high redshifts, and thus can be described reliably by the linear approximation. 
The halo abundance is basically strongly dependent on the underlying matter power spectrum, which can be approximated by either semi-analytical or fully numerical calculation.
The observed number of satellite galaxies in the Milky Way has been challenging to the standard $\Lambda$CDM cosmology because the number is much smaller than the values predicted theoretically. This discrepancy has been attributed usually to the dark matter property or the nonlinear baryonic physics, but the thermal inflation scenario can add an interesting alternative explanation for this discrepancy through its impact on the matter power spectrum.
While we focus mainly on the small-scale feature due to the thermal inflation scenario, it is also worthwhile to compare its predictions with those due to other plausible scenarios. Along this line, we will test the warm dark matter (WDM) scenario \citep{deVega:2009ku, deVega:2010yk} as well, where fluctuations at scales smaller than the free streaming length of WDM are suppressed.

The paper is organized as follows.
In \sref{sec:theory}, we review the evolution of perturbations under thermal inflation and show the thermal inflation %(curvature) 
	power spectrum.
In \sref{sec:massfun}, we show matter power spectra and halo mass functions at various epochs under thermal inflation and compare their results to those from WDM scenarios.
In \sref{sec:milkyway}, we investigate the small-scale effect of thermal inflation scenarios in terms of the abundance of satellite galaxies in Milky Way-sized galaxies.
In \sref{sec:21cm}, we quantify the expected power spectra of the 21-cm background fluctuations and provide forecasts their observability by the SKA.
We summarize our results in \sref{sec:summary}.

Throughout this paper, we adopt the following cosmological parameters by the 3rd-year {\em Planck} data \citep{Ade:2015}:
the matter density parameter $\Omega_{\rm m} = 0.3075$,
the cosmological constant parameter $\Omega_{\Lambda} = 0.6925$,
the baryon density parameter $\Omega_{\rm b} = 0.0486$,
the hubble parameter $h \equiv H_0 / (100 \kmsMpc) = 0.6774$,
the pivot scale $k_* = 0.05 \invMpc$,
the spectral index at the pivot scale $n_* = 0.97$,
and the amplitude of primordial power spectrum at the pivot scale $A_* = 2.14 \times 10^{-9}$.

\section{Power spectrum of thermal inflation scenario}\label{sec:theory}

\begin{figure}[tpb]
\plotone{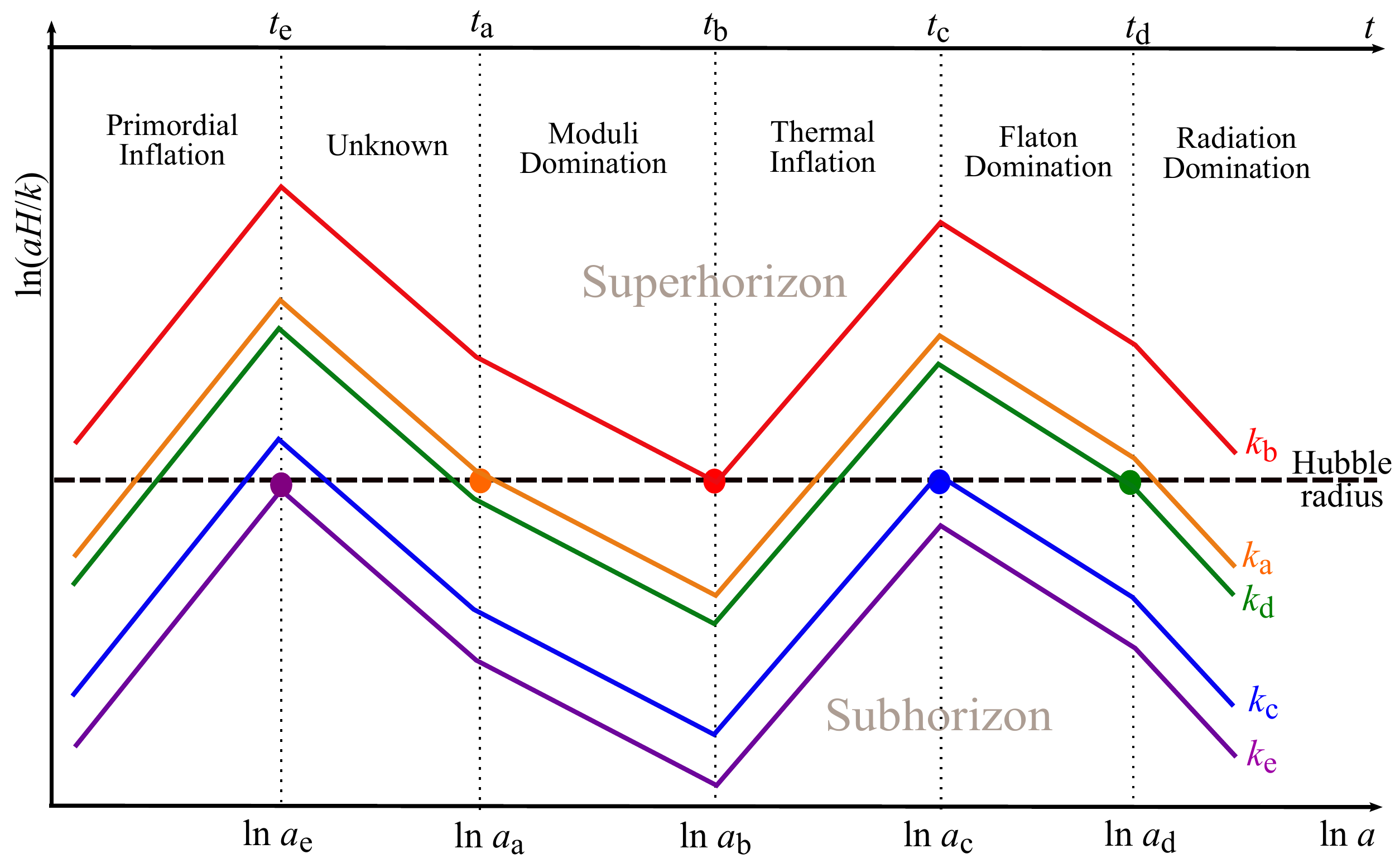}
\caption{
Characteristic scales of cosmology with thermal inflation.
There are five characteristic scales, $k_{\rm a}$, $k_{\rm b}$, $k_{\rm c}$, $k_{\rm d}$ and $k_{\rm e}$ corresponding to the comoving scale of the horizon at each of the era boundaries.
$k_{\rm b}$ is the largest, and hence the most observable scale.
} \label{fig:5scales}
\end{figure}

\fref{fig:5scales} shows the history of the Universe in the thermal inflation scenario \citep{Lyth:1995hj, Lyth:1995ka}.
In this scenario, the primordial inflation, ending at $t=t_{\rm e}$, is followed by thermal inflation after an unknown post-inflation epoch.  
Moduli matter starts dominating at $t=t_{\rm a}$ as the energy scale of the Universe drops below the moduli mass scale.
Then, thermal inflation commences at $t=t_{\rm b}$ as a second inflation and starts diluting the moduli matter. 
When the temperature of the Universe drops below the flaton's mass scale at $t=t_{\rm c}$, thermal inflation ends up with flaton matter domination. 
As the flaton rolls away from the origin, preheating process occurs and results in the usual radiation domination of BBN at $t=t_{\rm d}$.

\begin{figure}[tpb]
\plotone{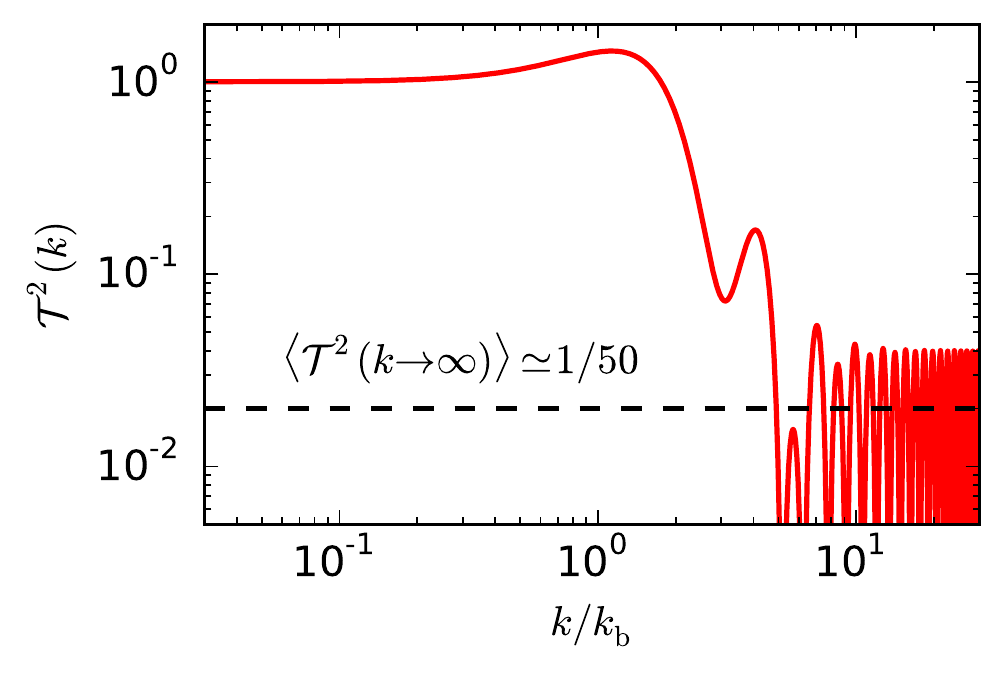}
\caption{Transfer function induced by thermal inflation, $\mathcal{T}_{\rm TI}^2(k)$, as a function of $k/\kb$ \citep{Hong:2015oqa}.}
\label{fig:transfer}
\end{figure}

We adopt the characteristic wavenumbers defined by \cite{Hong:2015oqa}, which allow a convenient description of the evolution of density perturbations during these eras. These are
$k_x \equiv \fn{a}{t_x} \fn{H}{t_x}$,
which correspond to the horizon at time $t_x$'s described above.
They are estimated as
\begin{align}\label{ti_scales}
k_{\rm a} & \simeq  7 \times 10^5 \invMpc 
\left(\frac{e^{20}}{e^{N_\mathrm{bc}}}\right) 
\left( \frac{m_\Phi}{\TeV} \right)^\frac{1}{3}
\left( \frac{T_{\rm reh}}{\GeV} \right)^\frac{1}{3} 
\\
k_{\rm b} &\simeq 3 \times 10^3 \invMpc
\left(\frac{e^{20}}{e^{N_\mathrm{bc}}}\right)
\left( \frac{T_{\rm reh}}{\GeV} \right)^\frac{1}{3}
\left( \frac{V_0^{1/4}}{10^7 \GeV} \right)^\frac{2}{3}
\\
k_{\rm c} & \simeq  3 \times 10^{11} \invMpc
\left( \frac{T_{\rm reh}}{\GeV} \right)^\frac{1}{3}
\left( \frac{V_0^{1/4}}{10^7 \GeV} \right)^\frac{2}{3} 
\\
k_{\rm d} & \simeq 2 \times 10^7 \invMpc
\left( \frac{T_{\rm reh}}{\GeV} \right) ,
\end{align}
where $N_{\rm bc}$ is the number of e-foldings during thermal inflation ($t_{\rm b} < t < t_{\rm c}$), the moduli mass scale is $m_{\Phi} \sim \TeV$, 
the vacuum potential energy during thermal inflation is $10^3 \GeV \ll V_0^{1/4} < 10^7 \GeV$, and the reheating temperature for the last radiation domination is $10^{-2}\GeV < T_{\rm reh} < 10^2 \GeV$.
$N_\mathrm{bc}$, which is a key parameter determining the impact of thermal inflation on the structure formation, should be greater than  $\sim 10 $ to resolve the moduli problem.
$N_\mathrm{bc} \lesssim 15$ is typical in single thermal inflation scenario where thermal inflation happens once throughout cosmological history. 
But $N_\mathrm{bc}$ can be even larger if we consider extended cases of multiple thermal inflation \citep{Lyth:1995ka,Felder:2007iz,Kim:2008yu,Choi:2012ye} and then $k_{\rm b}$ and $k_{\rm a}$ can become small enough to allow some observables. For example, $k_{\rm b} \simeq 20 \invMpc$ and $k_{\rm a} \simeq 5 \times 10^3 \invMpc$  when $N_{\rm bc} = 25$, which then can produce observable effects described in the following sections.
Because, e.g., the matter power spectrum in thermal inflation deviates significantly around and above $k_{\rm b}$ (see \fref{fig:transfer}) from those predicted by standard inflationary scenarios, we will take $\kb$ as the single parameter that fully quantifies the impact of thermal inflation on the cosmological structure formation.

The curvature power spectrum of the thermal inflation scenario $\fn{\mathcal{P}^{\rm TI}_{\mathcal{R}}}{k} $ can be expressed as
\begin{equation}\label{ps_TI}
\fn{\mathcal{P}^{\rm TI}_{\mathcal{R}}}{k}  = \fn{\mathcal{P}^{\rm pri}_{\mathcal{R}}}{k} \fn{\mathcal{T}_{\rm TI}^2}{k},
\end{equation}
where $\fn{\mathcal{P}^{\rm pri}_{\mathcal{R}}}{k}$ is the primordial curvature power spectrum that is from the primordial inflation, and
$\fn{\mathcal{T}_{\rm TI}}{k}$ is a transfer function that summarizes the evolution of curvature perturbation from $t_{\rm e}$ to $t_{\rm d}$. Afterwards, the observed matter power spectrum is determined by the standard transfer function $\fn{\mathcal{T}}{k;z}$ at redshift $z$ as
\begin{equation}\label{ps_obs}
\fn{P_{\rm m}}{k;z}  = \fn{\mathcal{P}^{\rm TI}_{\mathcal{R}}}{k} \fn{\mathcal{T}^2}{k;z},
\end{equation}
where $\fn{\mathcal{T}}{k;z}$ is determined by the evolution for $t\ge t_{\rm d}$ and can be obtained, in practice, from Boltzmann solvers such as the \textsc{camb} \citep{Lewis:1999,Howlett:2012}.

In simple inflation models, usually, the spectral index of primordial inflation is given by  
\begin{equation} \label{n_1overN}
\frac{d \ln \mathcal{P}^{\rm pri}_{\mathcal{R}} }{\d \ln k} \left( k \right)
= - \frac{c}{\mathcal{N}} 
+ \fn{\mathcal{O}}{ \frac{1}{\mathcal{N}^2} } ,
\end{equation}
where
\begin{equation}
\mathcal{N} \equiv
\left. \ln \frac{a_{\rm e} H_{\rm e}}{aH} \right|_{aH=k}
= \ln \frac{k_{\rm e}}{k}
\end{equation}
is the amount of inflation of a mode $k$ from its horizon exit ($aH = k$) to the end of inflation.
At the pivot scale of the matter power spectrum, or $k = k_{*} \simeq 0.001 h \invMpc$, 
\begin{align}
\left. \mathcal{P}^{\rm pri}_{\mathcal{R}} \right|_{k=k_*} & = A_{*}, \nonumber \\
\left. \frac{\d \ln \mathcal{P}^{\rm pri}_{\mathcal{R}} }{\d \ln k} \right|_{k=k_*} & =  n_* - 1 
\end{align}
(for impact on the spectral index of the primordial power spectrum, see e.g., \citealt{Dimopoulos:2016tzn, Dimopoulos:2016zhy, Dimopoulos:2017qqn, Cho:2017}).
Therefore, $\mathcal{P}^{\rm pri}$ is given by
\begin{equation} \label{ps_est}
\fn{\mathcal{P}^{\rm pri}_{\mathcal{R}}}{k} 
\simeq A_* \left( 1 - \frac{1}{\mathcal{N}_*} \ln \frac{k}{k_*} \right)^{ (1-n_*) \mathcal{N}_* }  ,
\end{equation}
with 
\begin{equation}  \label{N*}
\mathcal{N}_* \equiv \ln \frac{k_{\rm e}}{k_*} ~.
\end{equation}
Note that  $\mathcal{N}_*$ is somewhat uncertain, mainly due to ambiguities in the equation of state $w=P/\rho$ between $t_{\rm e}$ and $t_{\rm a}$ and in the scale of thermal inflation potential energy.  
Assuming that the era between $t_{\rm e}$ and $t_{\rm a}$ is governed by a single equation of state with $0 \leq w \leq 1/3$, and the scale of thermal inflation is between $10^5 \GeV$ and $10^9 \GeV$, $\mathcal{N}_*$ lies in the following range \citep{Cho:2017}:
\begin{equation}\label{NTIrange}
13 + \ln \left( \frac{k_{\rm b}}{10^3 \invMpc} \right)
\lesssim \mathcal{N}_* \lesssim 
32 + \ln \left( \frac{k_{\rm b}}{10^3 \invMpc} \right)  .
\end{equation}

\cite{Hong:2015oqa} showed that $\fn{\mathcal{T}_{\rm TI}}{k}$ is, in fact, a simple function of $k / \kb$ (see \fref{fig:transfer}).
Its asymptotic form is given by
\begin{equation}\label{kappazero}
\fn{\mathcal{T}_{\rm TI}}{\frac{k}{\kb}} \to
\begin{cases}
\begin{displaystyle} 1 + \nu_0 \left( \frac{k}{\kb} \right)^2 + \mathcal{O}\left[\left( \frac{k}{\kb} \right)^4\right] \end{displaystyle} & \begin{displaystyle} \textrm{ as } \frac{k}{\kb}  \to 0 \end{displaystyle} \\
\begin{displaystyle} - \frac{1}{5} \cos \left[ \nu_1 \left( \frac{k}{\kb} \right) \right] + \mathcal{O}\left[\left( \frac{k}{\kb} \right)^{-n} \right] \end{displaystyle} & \begin{displaystyle}\textrm{ as } \frac{k}{\kb} \to \infty \end{displaystyle}
\end{cases}
\end{equation}
where $\nu_0 \simeq 0.3622$ and $\nu_1 \simeq 2.2258$.
$\fn{\mathcal{T}_{\rm TI}^2}{k/\kb}$ is close to 1 at $k \lesssim \kb$
but has a substantial enhancement at $k = 1.1 \kb$ by  $\sim46\%$, and for larger $k$ the suppression becomes significant. $\fn{\mathcal{T}_{\rm TI}^2}{k/\kb}\simeq 1/14$ at $k = 3.1 \kb$, and $\fn{\mathcal{T}_{\rm TI}^2}{k/\kb}$ oscillates around $1/50$ at $k \gg 3 \kb$.

Henceforth, we take the term ``standard $\Lambda$CDM scenario'' as the case with a pure power-law primordial spectrum without a running spectral index,
\begin{equation}\label{LCDM}
\fn{\mathcal{P}^{\rm pri, \Lambda CDM}_{\mathcal{R}}}{k}  = A_* \left( \frac{k}{k_*} \right)^{n_*-1}~.
\end{equation}
Compared to the standard $\Lambda$CDM scenario, the curvature power spectrum of the thermal inflation scenario shows the following features: (1) a modest change of spectral index over $k$ by the amount of inflation $\mathcal{N}_*$ (\eqnp{ps_est}), (2) a boost of $\pk$ at $k\simeq \kb$, and (3) a strong suppression of $\pk$  at $k \gtrsim 3\kb$ (see \fref{fig:transfer}).

\section{Matter Power Spectrum and Halo Distribution}\label{sec:massfun}

In this section, we study how thermal inflation affects the matter and halo distribution of the Universe.
As shown in the previous section, the most noticeable difference between thermal inflation and standard $\Lambda$CDM scenarios is the suppression of the matter power spectrum at $k \gtrsim 3 \kb$.
This would also cause a suppression of the number of low-mass halos whose mass is less than 
\begin{equation}
\Msupp{TI} \simeq {4\pi \over 3} \bar{\rho}_{\rm m} \left({2\pi \over 3 \kb}\right)^3 \sim 10^{11} \msun \left({\kb \over \invMpc}\right)^{-3}  ,
\end{equation}
where $\bar{\rho}_{\rm m}$ is the mean matter density at $z = 0$.

The suppression of the matter power spectrum at $k \gtrsim 3 \kb$ could be used to constrain a valid range of $\kb$.
Since the primordial power spectrum reconstructed from CMB observations does not have such a huge suppression at $k \lesssim 1 \invMpc$, $\kb$ should be larger than $1 \invMpc$ \citep{Ade:2015, Hong:2015oqa}.
On the other hand, if $\kb$ is much larger than $10^2 \invMpc$, then low-mass halos suffering the suppression would be smaller than $\Msupp{TI} \lesssim 10^5 \msun$, which is roughly the Jeans mass under the thermal condition of the intergalactic medium at high redshifts \citep[e.g.,][]{Naoz:2009}.
Therefore, only the thermal inflation models with $\kb \lesssim 10^2 \invMpc$ could give observable signatures in the matter power spectrum and the halo population\footnote{In principle, dark matter dominated halos at mass scales smaller than the Jeans mass can exist. These can affect e.g. the gamma-ray background from dark matter annihilation (e.g. \citealp{Ahn:2004yd}). However, we defer a study of such small-scale, indirect observables caused by thermal inflation.}.
From now on, therefore, we focus on thermal inflation scenarios with $1 \invMpc \leq \kb \leq 10^2 \invMpc$.

Note that a similar suppression of the number of low-mass halos is expected in the WDM scenarios---in this case, the suppression of matter power spectrum by a factor of 50 occurs at $k = 2.5 \khalf$, where $\khalf$ is defined in \eq{k12}.
\cite{Destri:2013} shows that the transfer functions from various WDM scenarios can be generalized as
\begin{equation}
\fn{\mathcal{T}_{\rm WDM}^2}{k} \approx \left[1 + 0.167 \left(\frac{k}{\khalf} \right)^{2.304} \right]^{-4.478}  ,
\end{equation}
in which
\begin{equation}\label{k12}
\khalf \approx 6.72\invMpch \left( \frac{\mFD}{\keV} \right)^{1.12}
\end{equation}
is the characteristic scale where the matter power spectrum is suppressed by half, and $\mFD$ is the WDM particle mass for fermion decoupling at thermal equilibrium.
Therefore, the suppression mass scale for the WDM scenario corresponding to  $\Msupp{TI}$ in the thermal inflation scenario can be written as
\begin{equation}
\Msupp{WDM} \simeq {4\pi \over 3} \bar{\rho}_{\rm m} \left({2\pi \over 2.5 \khalf}\right)^3 \sim 3 \times 10^{10} \msun \left({\mFD \over \keV}\right)^{-3}  .
\end{equation}

\begin{figure}[htb]
\plotone{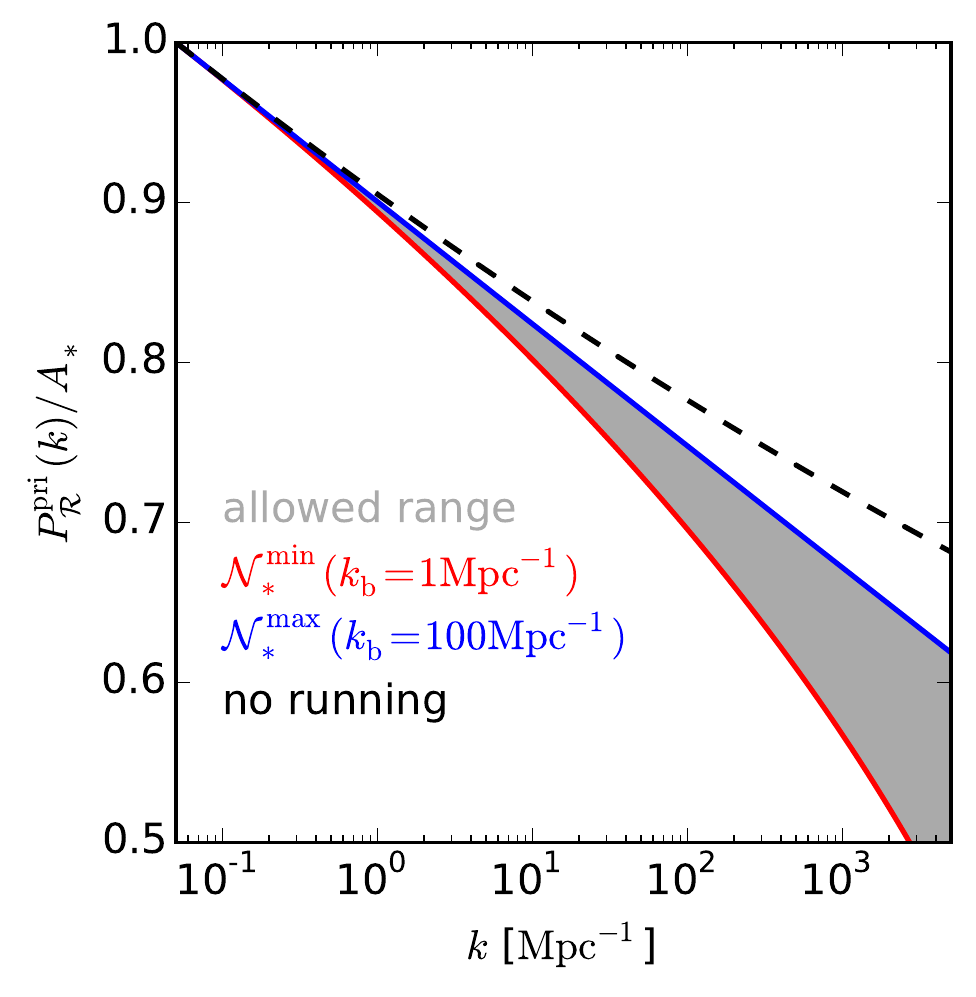}
\caption{The allowed range of primordial power spectra $\mathcal{P}_{\mathcal{R}}^{\rm pri}$ in thermal inflation scenario with $1 \invMpc \leq k_{\rm b} \leq 10^2 \invMpc$, normalized with the amplitude at the pivot scale $k_* = 0.05 \invMpc$.
Black dash: Standard $\Lambda$CDM scenario without spectral index running.}\label{fig:running}
\end{figure}

\begin{figure}[htb]
\epsscale{1.15}
\plottwo{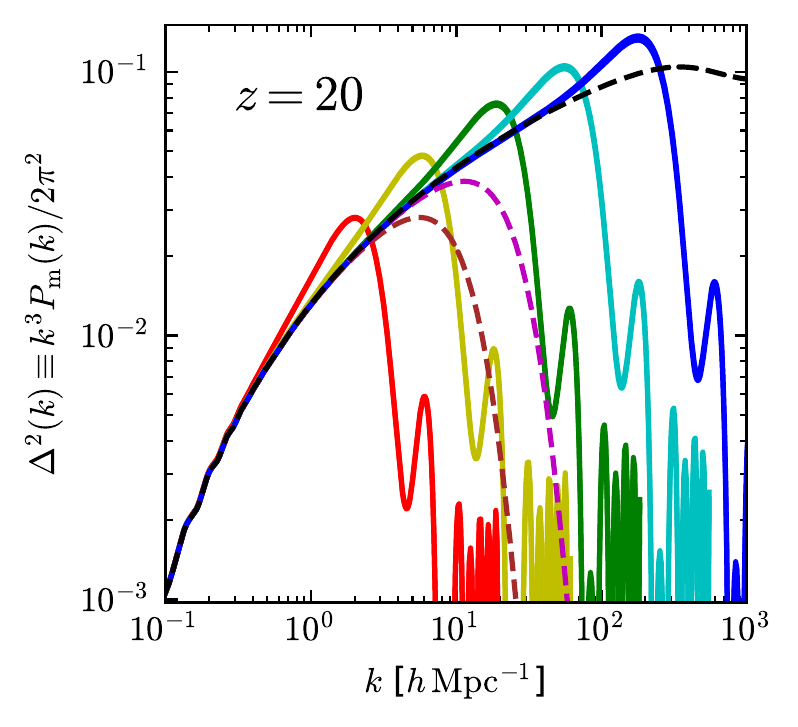}{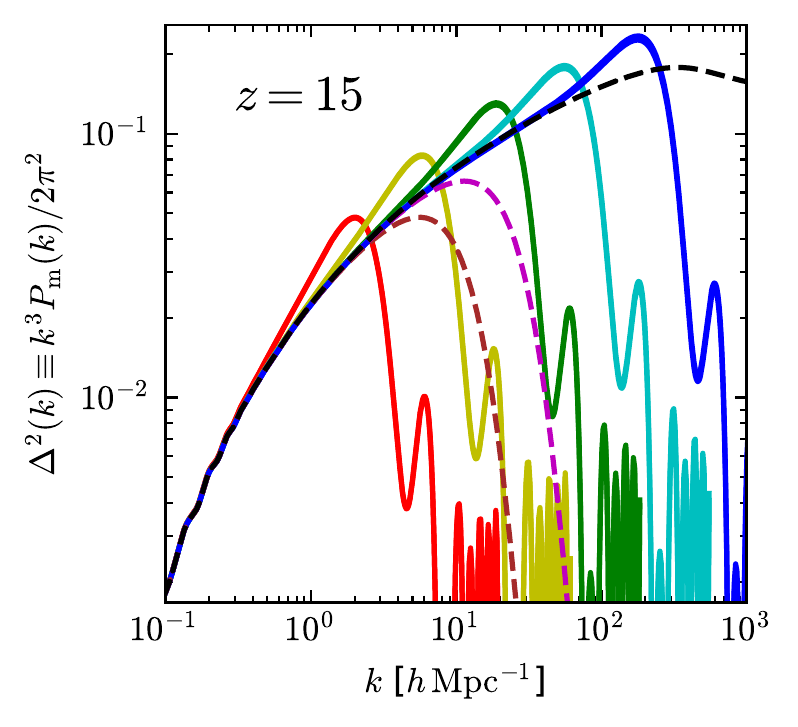}
\plottwo{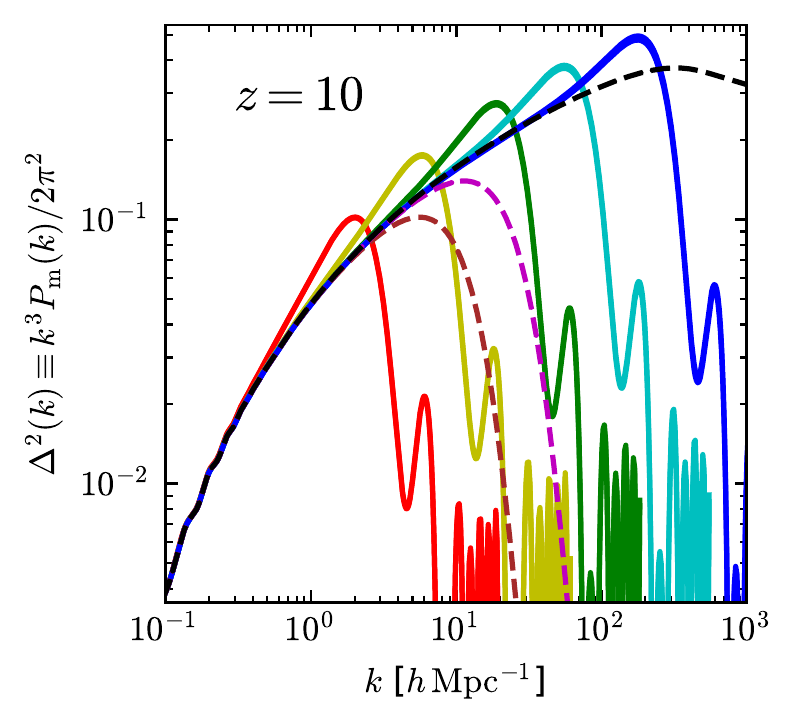}{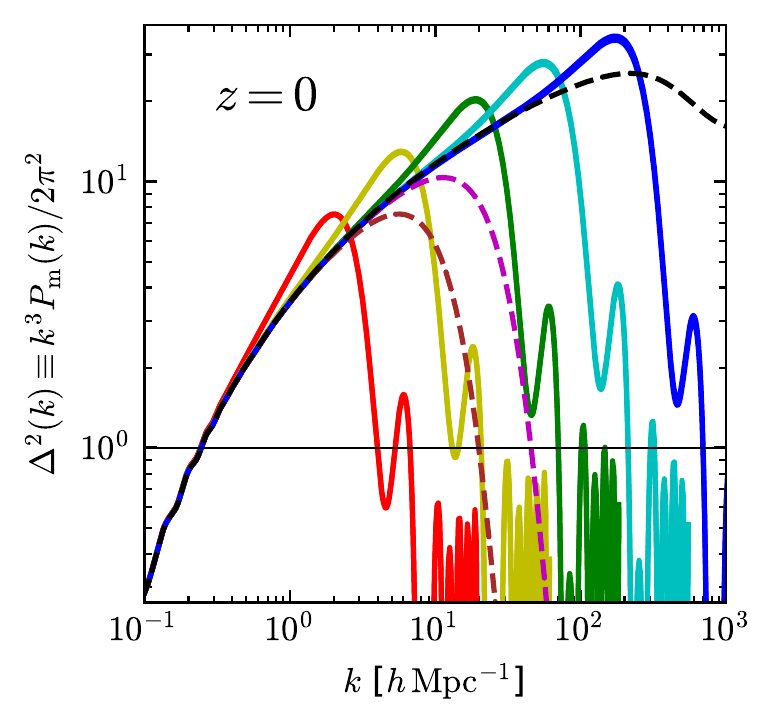}
\caption{Evolution of matter power spectra $\Delta^2(k) \equiv k^3 \pk/2\pi^2$ for thermal inflation and warm dark matter scenarios from $z = 20$ to $0$.
Solid lines: thermal inflation scenarios with $\kb = ${\color{red} $1 \invMpc$ (red)}, {\color{Goldenrod} $3 \invMpc$ (yellow)}, {\color{OliveGreen} $10 \invMpc$ (green)}, {\color{cyan} $30 \invMpc$ (cyan)}, and {\color{blue} $100 \invMpc$ (blue)}.
The change of power spectra from the choice of different $\mathcal{N}_*$ is negligible.
Dashed lines: WDM scenarios with $\mFD =${\color{brown} $1 \keV$ (brown)} and {\color{magenta} $2 \keV$ (magenta)}. 
Black dash: standard $\Lambda$CDM scenario.}\label{fig:matterpower}
\end{figure}

\fref{fig:running} shows how thermal inflation affects running of the spectral index of $\fn{\mathcal{P}_{\mathcal{R}}^{\rm pri}}{k}$ for $1 \invMpc \leq \kb \leq 10^2 \invMpc$.
In a thermal inflation scenario, the spectral index of the power spectrum slightly changes over $k$, and the suppression is stronger in smaller $\kb$.
Suppression of $\fn{\mathcal{P}^{\rm pri, \Lambda CDM}_{\mathcal{R}}}{k}$ due only to this running spectral index remains less than 1\% at $k = 1\invMpc$ and $\sim 7-27\%$ at $k = 10^3 \invMpc$. For $k\lesssim \kb$, this mild suppression is completely offset by the enhancement in $\fn{\mathcal{T}_{\rm TI}^2}{k}$ around $k=\kb$, as seen in \fref{fig:matterpower}.

\fref{fig:matterpower} shows the evolution of matter power spectra for both thermal inflation and WDM scenarios with various $\kb$'s and $\mFD$'s from $z = 20$ to $0$.
We calculate $\pk$ by using a modified \textsc{camb} code with $\fn{\mathcal{P}_{\mathcal{R}}^{\rm TI}}{k}$'s as inputs.
Throughout these redshifts, as noted in \sref{sec:theory}, $\pk$ of the thermal inflation scenario is substantially boosted from that of the standard $\Lambda$CDM scenario at $\kb \lesssim k \lesssim 3 \kb$, and then is strongly suppressed at $\kb \gtrsim 3\kb$ (see \fref{fig:transfer}).
At $z = 0$, the immature growth of structures (i.e. $\Delta^2 (k) \equiv k^3 \pk / 2\pi^2 \lesssim 1$) occurs at $k > k_{\rm supp}$, where
\begin{equation}
k_{\rm supp} \simeq
\begin{cases}
5 \kb & \textrm{ for thermal inflation scenarios} \\
\begin{displaystyle} 3 \khalf \simeq 20 \invMpch \left( \frac{\mFD}{\keV} \right) \end{displaystyle} & \textrm{ for WDM scenarios.}
\end{cases}
\end{equation}
Note that, for a fixed $\kb$ or $N_{\rm bc}$, the difference in $\pk$ caused by varying $\mathcal{N}_*$ (\eqnp{NTIrange}) is negligible in \fref{fig:matterpower}, affecting only the oscillating regime at $k\gg \kb$.

\begin{figure}[htb]
\plotone{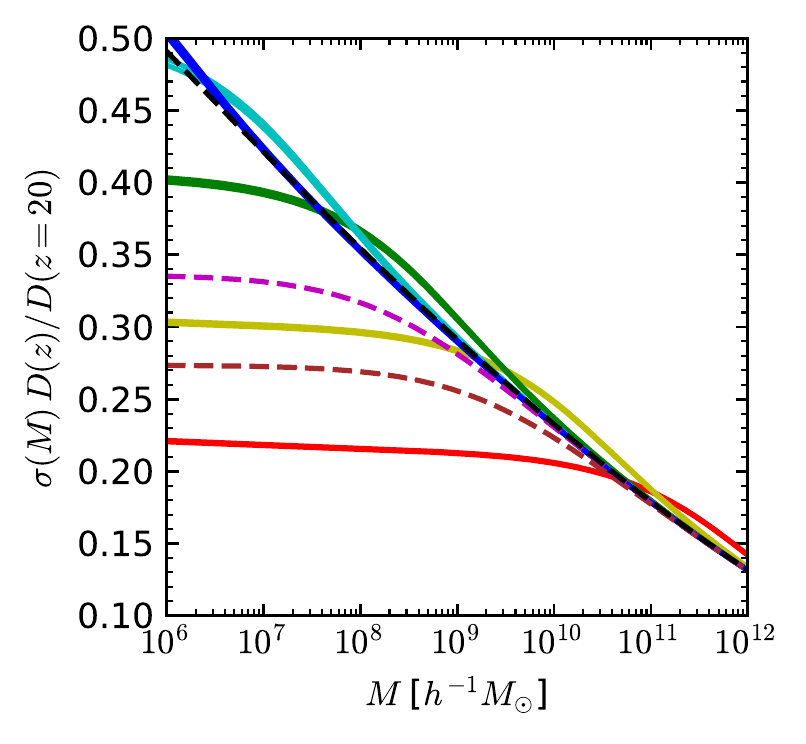}
\caption{The standard deviation of the matter density field smoothed by a window function with length scale, $\fn{R}{M} \equiv ({3 M / 4\pi \bar{\rho}_{\rm m}})^{1/3}$, for a given mass scale $M$, for thermal inflation and WDM scenarios.
Color codes and line styles are the same as \fref{fig:matterpower}.}\label{fig:density}
\end{figure}

The variance of the matter density field smoothed by a window function with length scale $\fn{R}{M} \equiv ({3 M / 4 \pi \bar{\rho}_{\rm m}})^{1/3}$ for a given mass scale $M$ is given by
\begin{equation}\label{eq:sigmaM}
\fn{\sigma^2}{M} \equiv \left\langle\fn{\delta^2}{{\bf x};\fn{R}{M}}\right\rangle =\frac{1}{2 \pi^2} \int^{\infty}_0 \d{k} k^2 \fn{P_{\rm m}}{k} \fn{W^2}{ k \fn{R}{M}} ,
\end{equation}
where the filtered density field given by the convolution with the window function:
\begin{equation}\label{eq:delta_filtered}
\fn{\delta}{{\bf x};\fn{R}{M}}=\int \delta({\bf x'})F(|{\bf x}-{\bf x'}|;\fn{R}{M})%,
\end{equation}
is an indicator of whether halos with the corresponding mass $M$ have collapsed or not.
Here 
\begin{equation}
\delta({\bf x})=\frac{\rho(x)-\bar{\rho}_{\rm m}}{\bar{\rho}_{\rm m}}
\end{equation}
is the matter overdensity,
\begin{equation}
F(|{\bf x}-{\bf x'}|;\fn{R}{M})=
\begin{cases} 
\begin{displaystyle} \frac{3}{4 \pi \fn{R^3}{M}} \end{displaystyle} & \textrm{if } |{\bf x} - {\bf x'}| \le \fn{R}{M}\\
0 & \textrm{otherwise}
\end{cases}
\, ,
\end{equation}
is the commonly used spherical top-hat window function, and 
\begin{equation}
\fn{W}{x} \equiv \frac{2 \left( \sin x - x \cos x \right) }{x^3}
\end{equation}
is the corresponding $k$-space window function \citep[e.g.,][]{Spergel:2006}.

\fref{fig:density} shows $\fn{\sigma}{M}$ for thermal inflation and WDM scenarios. 
In both scenarios, $\fn{\sigma}{M}$ becomes clearly lower than the standard $\Lambda$CDM scenario with a gentle slope at $M \ll \Msupp{}$, because of the suppression of matter power spectrum at $k \gg \kb$ or $k \gg \khalf$.
On the other hand, around $k \sim \kb$, thermal inflation cases show the boost from the standard $\Lambda$CDM value in $\fn{\sigma}{M}$ at $\Msupp{TI} \lesssim M \lesssim 10^2 \Msupp{TI}$. This is again due to the enhancement in $\pk$ around $k=\kb$.
Finally, for massive halos with $M \gtrsim 10^2 \Msupp{TI}$, all models converge to the standard $\Lambda$CDM prediction.

\begin{figure}[htb]
\epsscale{1.15}
\plottwo{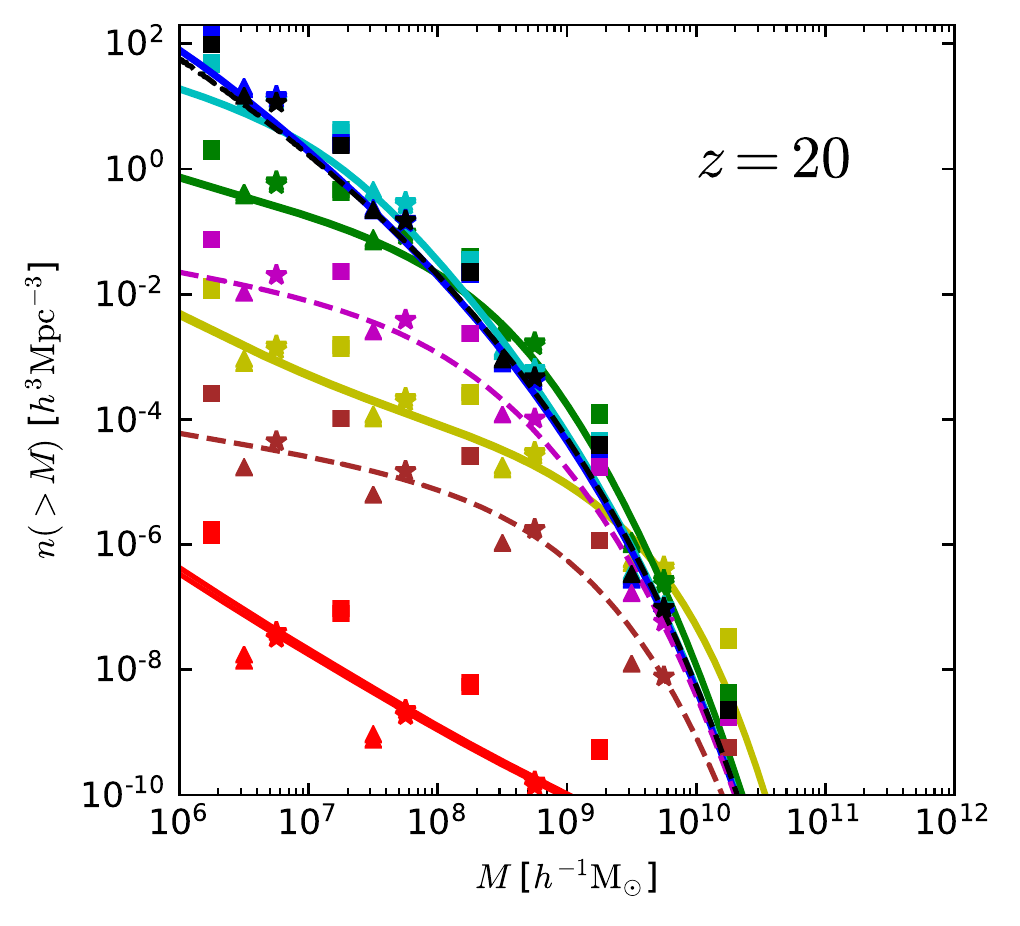}{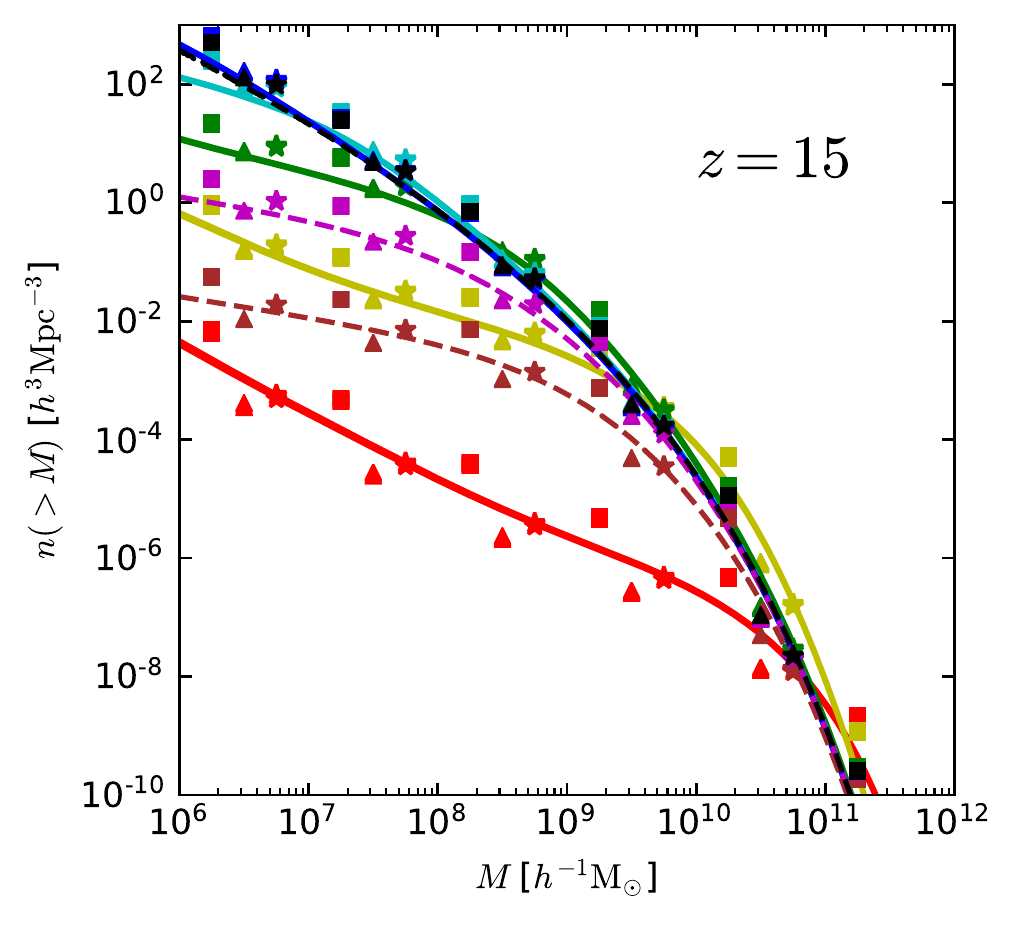}
\plottwo{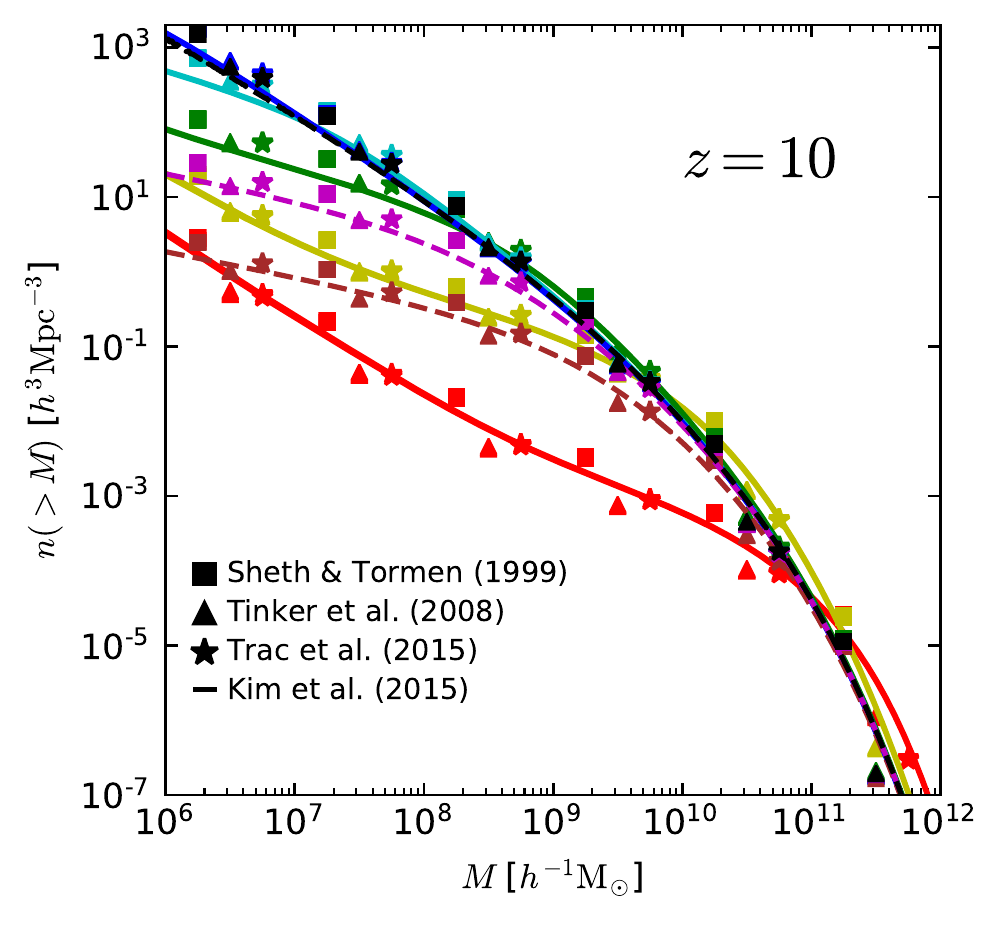}{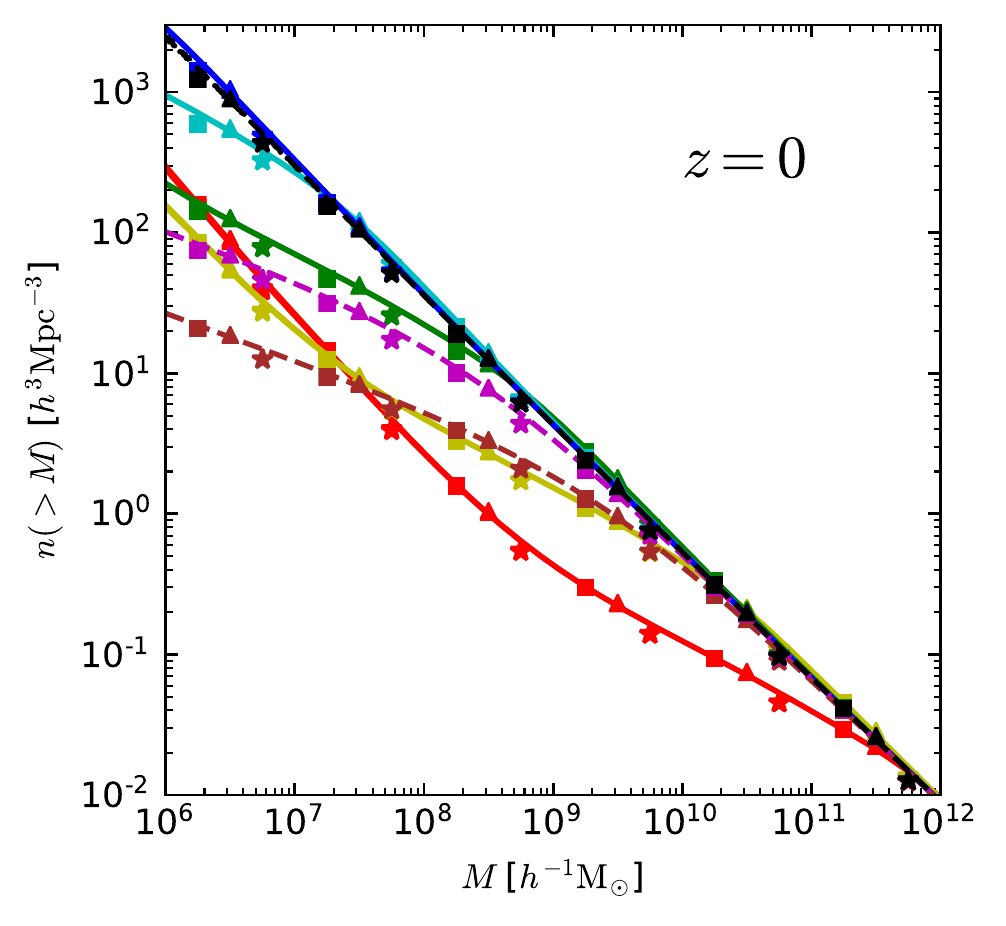}
\caption{Evolution of global halo mass functions $n(> M)$ for thermal inflation and WDM scenarios from $z = 20$ to $0$, calculated by the various mass function models (symbols).
Lines: the mass function described in \citet{Kim:2015}.
Color codes and line styles are the same as \fref{fig:matterpower} and \ref{fig:density}.}\label{fig:massfun}
\end{figure}

\fref{fig:massfun} shows global halo mass functions for thermal inflation and WDM scenarios from $z = 20$ to $0$, calculated by adopting the four mass functions described in \citet{Sheth:1999}, \citet{Tinker:2008}, \citet{Trac:2015}, and \citet{Kim:2015}.
As expected, the suppression of $\pk$ at $k \gtrsim 3 \kb$ induces the suppression of low-mass halo population with $M \lesssim \Msupp{TI}$.
Compared to the standard $\Lambda$CDM scenario, the suppression of low-mass halo population with mass $M \lesssim 10^7 \msunh$ could reach a factor of $\sim 10^2$ in the case of $\kb = 1 \invMpc$.
On the other hand, thermal inflation scenario with $\kb \gg 10 \invMpc$ has only a negligible amount of suppression or even a slight enhancement of low-mass halo population, because with such high $k_{\rm b}$ the enhancement of $\pk$ around $k \simeq \kb$ becomes the major factor.

While suppression of low-mass halo population is significant for thermal inflation models with $\kb \lesssim 10 \invMpc$, the enhancement of massive halo population would be hard to detect because this effect is too mild at low redshifts to beat other uncertainties such as the galaxy bias. Although the enhancement of massive halo population with $M \gtrsim 10^8 \msunh$ is more pronounced at high redshifts when $\kb \lesssim 10 \invMpc$, observing these halos is extremely difficult due to the low brightness of the associated galaxies, and the sample variance becomes high for rare objects. Nevertheless, high-sensitivity galaxy probes such as the {\em James Webb Space Telescope} (JWST; \citealt{JWST:2006}) and the {\em Extremely Large Telescope-Multi-AO Imaging Camera for Deep Observations} (ELT-MICADO; \citealt{MICADO:2016}) may shed some light on the nature of these objects.

\section{Local properties: abundance of satellite galaxies inside the Milky Way}\label{sec:milkyway}

The over-abundance of subhalos inside the numerically simulated Milky Way environment against the actual, observed number of the Milky Way's subhalos had once been regarded as a prime challenge to the $\Lambda$CDM paradigm \citep{Moore:1999}. 
There have been several possible explanations for the above question, such as 
(1) the actual dark matter property may be fundamentally different from that of the pure CDM at the scale of subhalo formation, such that the halo formation is suppressed by free-streaming or self-interaction \citep{Rocha:2013};
(2) previously neglected baryonic physics, such as the boosted Jeans-mass filtering by photo-heating from the cosmic reionization, may be strong enough to suppress star formation inside these halos and thus can prevent some of the halos from being observed \citep[e.g.][]{Susa:2004}; and
(3) somewhat poorly calculated dynamics of satellite halos, such as the tidal stripping \citep[e.g.][]{Kravtsov:2004} and the supernova feedback \citep[e.g.][]{Font:2011}, may be responsible for suppressing the halo formation. 

Even though the latter two explanations may be appealing in a sense that we need not give up the $\Lambda$CDM paradigm, they are still inconclusive because we cannot directly probe the evolution of these halos. 
On the other hand, the thermal inflation scenario has potential to explain the above problem, by reducing the abundance of subhalos themselves from the reduced matter power spectrum at small scales. Recently, this problem has evolved into the discrepancy in the internal dynamical properties of the most massive satellites \citep{B-K:2012}, and thus resolutions regarding the dark matter halo itself (including ours) are more favored than the ones regarding only the baryon physics. 

To study the satellite abundance in the Milky Way-sized galaxies in thermal inflation scenario, we performed a series of simplified cosmological $N$-body simulations for various thermal inflation and WDM scenarios by using the \textsc{pinocchio} code \citep{Monaco:2002} which adopts a semi-numerical approach based on the 2nd-order Lagrangian perturbation theory and the extended Press-Schechter formalism. 
Instead of finding dark matter (DM) halos (``halos'' hereafter) and their merger histories after finishing the evolution of DM particles in multiple timesteps, \textsc{pinocchio} directly generates halo merger trees by estimating the time when the matter at each grid point encounters the spherical collapse. This way, one can simulate the formation and evolution of cosmological halos much more quickly than the usual simulations using $N$-body gravity solvers, and also easily track the formation and evolution of satellites inside a host halo through its halo merger tree.

Each simulation is performed in a periodic, cubic box with a comoving volume $V_{\rm box} = (12.8 \Mpch)^3$ and $256^3$ grid points, which is able to cover $\pk$ up to the Nyquist frequency $k_{\rm Nyquest}$ at $ 62.8 \invMpch$. 
Halos are defined as those having more than 30 DM ``particles'' (i.e., having the mass greater than $3 \times 10^8 \msunh$), which are actually Lagrangian grid points in \textsc{pinocchio}.
Instead of running a constrained-realization simulation, which we defer to future work, we performed 10 different simulations based on 10 different realizations of the initial conditions (with varying random seeds) for each inflationary scenario.  We then sampled halos having the virial mass of the Milky Way. In each of the halos selected this way, we then sampled satellite galaxies. 

We then take a few more steps beyond just generating the DM halo catalogs and the merger trees by \textsc{pinocchio}.
From the synthesized merger tree for halos with  $M > 3 \times 10^8 \msunh$, we assign mock galaxies by using a one-to-one correspondence method described in \citet{Hong:2016}.
This method was used to generate a mock galaxy catalog of a large cosmological $N$-body simulation \citep{Kim:2015}, which has been tested against actual galaxy surveys through various analyses, such as the galaxy two-point correlation function \citep{Li:2016} and the two-dimensional topology \citep{Appleby:2017}.
In this method, each halo in the merging history of a given host halo becomes a galaxy candidate in the halo.
The most massive halo in the merging history corresponds to the host galaxy.
Other halos correspond to the satellite galaxy ``candidates'', and only those which are not tidally disrupted are marked as the surviving satellite galaxies.
In order to implement the tidal disruption of satellites, we choose surviving satellites of mass $M_{\rm sat}$ among the candidates if the time they took after their first merger into the tree of a host of mass $M_{\rm host}$ is larger than the tidal disruption time given by
\begin{equation}
t_{\rm disrupt} = \frac{(0.94 \epsilon^{0.60} + 0.60)/0.86}{\ln \left[1 + (M_{\rm host}/M_{\rm sat}) \right]}
\frac{M_{\rm host}}{M_{\rm sat}} \frac{R_{\rm vir}}{V_{\rm vir}},
\end{equation}
which is an empirical fit by \citet{Jiang:2008}.
Here $R_{\rm vir}$ and $V_{\rm vir}$ are the virial radius and the circular velocity at the virial radius of the host halo. 
$\epsilon$ is the ellipticity of the satellite's orbit, and we set it to a universal value 0.5, which is roughly the mean value found in other simulations.
Note that a merger event of two halos is not identical to that of two galaxies: e.g. two halos, each containing a single galaxy, may merge into one halo but two galaxies may still reside inside the halo as individual entities.

\begin{figure}[htb]
\epsscale{1.15}
\plottwo{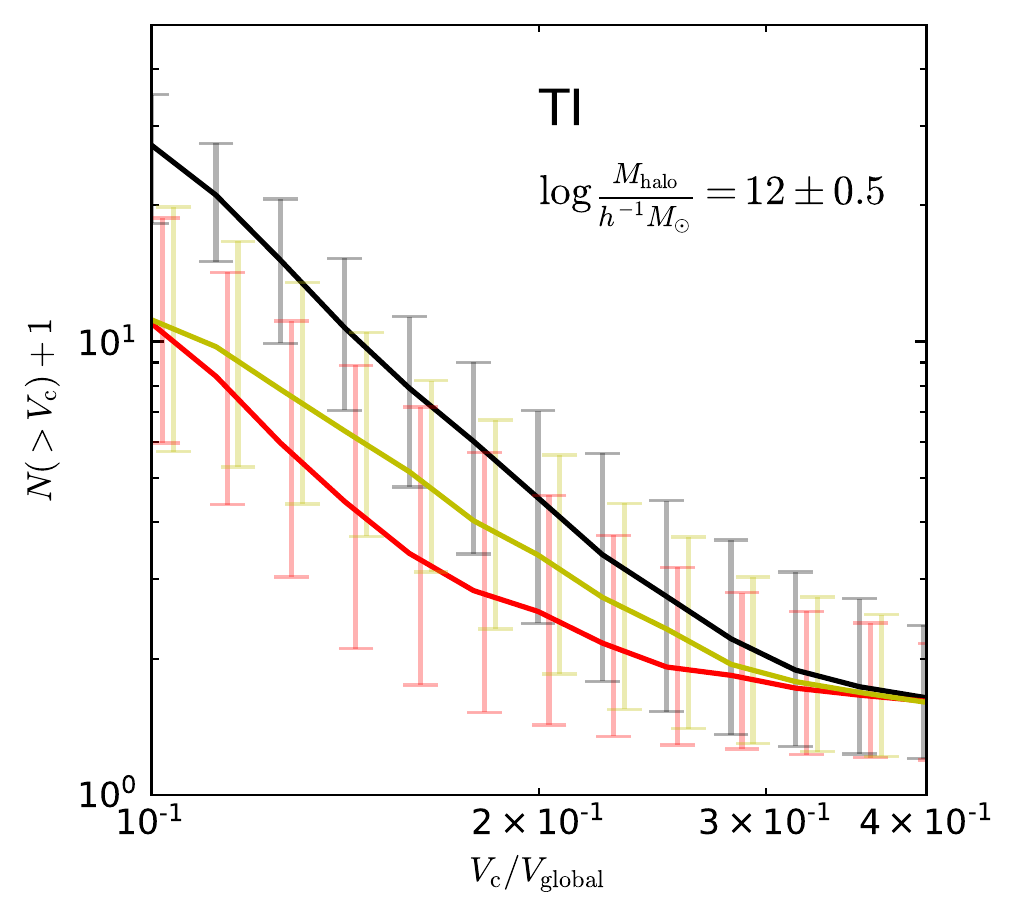}{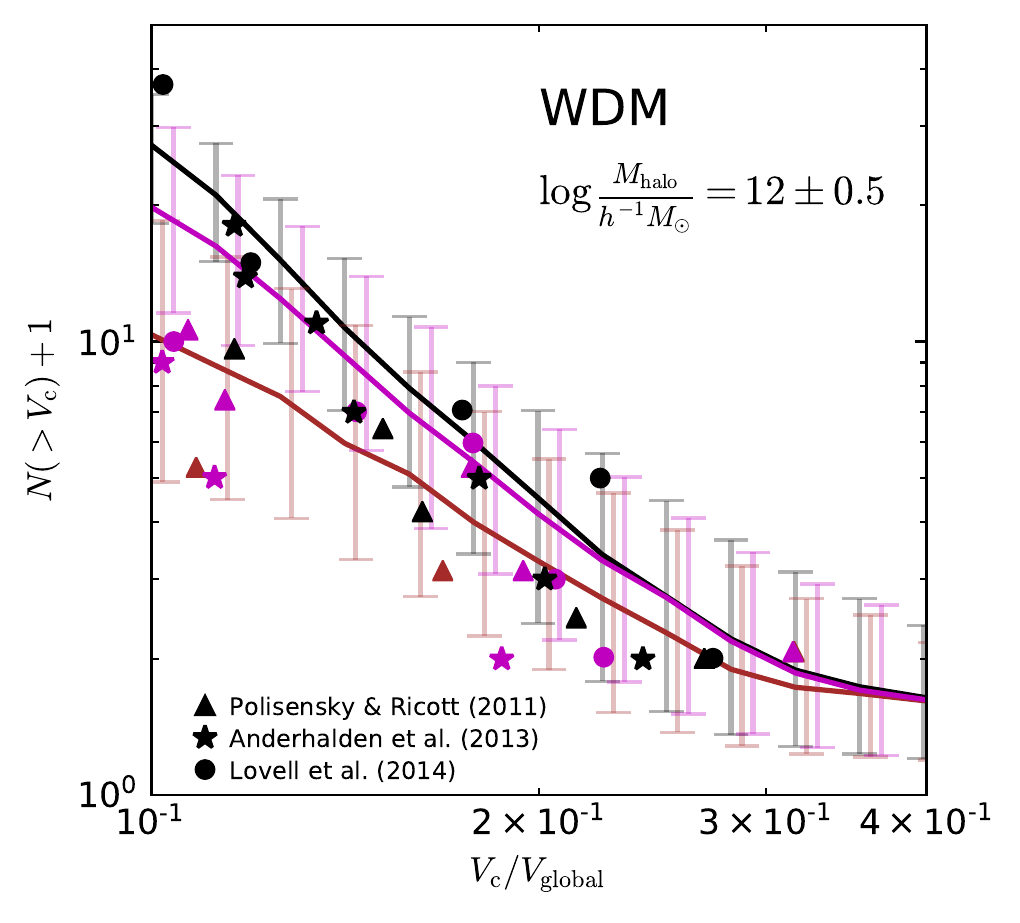}
\caption{The cumulative number of satellite galaxies of simulated Milky Way-size halos, $N(>V_{\rm c})$, for thermal inflation scenarios (left panel) with $\kb =$ {\color{red} $1\invMpc$ (red)}, {\color{Goldenrod} $3\invMpc$ (yellow)}, WDM scenarios (right panel) with $\mFD =$ {\color{brown} $1\keV$ (brown)}, {\color{magenta} $2\keV$ (magenta)}, and the standard $\Lambda$CDM scenario (black) at $z = 0$.
$V_{\rm global}$ and $V_{\rm c}$ are the circular velocities of the Milky Way and one of its satellite galaxies indicating its mass ($V_{\rm c} \propto M^{1/3}$).
Line and error bar show the median and $1\sigma$ deviation.
Symbols: results from previous studies on satellite abundance in WDM scenarios, by adopting $V_{\rm global} = 218 {\rm km/s}$ \citep{Bovy:2012}.}\label{fig:velfun_sim}
\end{figure}

\fref{fig:velfun_sim} shows the abundance of satellite galaxies in simulated Milky Way-size halos, with $\log (M_{\rm halo} / \msunh) = 12 \pm 0.5$.
In our simulations, 321 halos are found within the above mass range in the standard $\Lambda$CDM in total.
In this figure, instead of directly using DM mass, we plot the cumulative number of satellite galaxies as a function of the circular velocity of satellites ($V_{\rm c}$) in the unit of the circular velocity of the host ($V_{\rm global}$), by assuming the virial relation $V \propto M^{1/3}$ at present.
Note that, in all cases, the variance in the satellite abundance is large, such that the difference between the medium and $1\sigma$ deviation is greater than $0.2~{\rm dex}$.

We also show $N(>V_{\rm c})$ of the standard scenarios and WDM scenarios from \citet{Polisensky:2011}, \citet{Anderhalden:2013}, and \citet{Lovell:2014} as a comparison.
Note that there exists large deviation among the satellite abundance from previous studies, while most of them fit within $1\sigma$ deviation from the medium in our result.

Both thermal inflation scenarios show a deficit in the satellite abundance from that of the standard $\Lambda$CDM model. 
In the thermal inflation scenario with $\kb = 1 \invMpc$, the abundance of satellite galaxies can be clearly distinguished to that of the standard $\Lambda$CDM scenario for small satellites with $V_{\rm c} \lesssim 0.2 V_{\rm global}$ (i.e., $M_{\rm sat} \lesssim 10^{10} \msunh$).
At $V_{\rm c} = 0.1 V_{\rm global}$ (i.e., $M_{\rm sat} \simeq 10^9 \msunh$), the number of satellites is expected to be $1/3$ of those from the standard $\Lambda$CDM scenario on average. 
The thermal inflation scenario with $\kb = 3 \invMpc$ has a weaker deviation, with more overlap with the standard $\Lambda$CDM scenario at $1\sigma$ level.
However, the abundance of satellite galaxies at $V_{\rm c} \lesssim 0.1 V_{\rm global}$ (i.e., $M_{\rm sat} \lesssim 10^9 \msunh$) is almost identical to the scenario with $\kb = 1 \invMpc$. 
Note that the mass scale of satellite galaxies that abundance of satellite galaxies from $\kb = 1 \invMpc$ and $3 \invMpc$ becomes similar is $M_{\rm sat} \sim 10^9 \msunh$, which is similar to the halo mass scale where the mass functions from those two $\kb$'s meet. The distinction between the two models is the most prominent at $V_{\rm c}\simeq 1.7 V_{\rm global}$ when the average values are compared.

The WDM scenario with $\mFD = 1 \keV$ has similar satellite galaxy abundance to the thermal inflation scenario with $\kb = 3 \invMpc$ on average, while the WDM scenario with $\mFD = 1 \keV$ has larger abundance than the thermal inflation scenario with $\kb = 3 \invMpc$. Because of this, it would be difficult to distinguish between the WDM scenario with $\mFD = 1 \keV$ and the standard $\Lambda$CDM scenario purely based on the number of Milky Way satellites. There also seems to exist a discrepancy between the thermal inflation scenario with $\kb = 1 \invMpc$ and the WDM scenario with $\mFD = 1 \keV$ at $V_{\rm c}\simeq 1.7 V_{\rm global}$, but to be conclusive we need many more simulation data.

One caveat of our analysis is that this is based not on a constrained realization of the Local Group but on a series of mean-density realizations, and it is still to be answered whether the wide variance we observe in our suite of realizations will shrink in one or more constrained realizations.

\section{21-cm Power Spectrum}\label{sec:21cm}
The 21-cm line from the neutral hydrogen atoms can be measured against the CMB, which can be used to probe the distribution of the baryonic gas. 
The signal is quantified by the differential brightness temperature $\dTb$, defined by
\begin{equation}
\dTb = \frac{T_{S}-T_{\rm CMB}}{1+z} (1-e^{-\tau}),
\label{eq:dTb_def}
\end{equation}
where $T_{S}$ is the spin temperature of the singlet-triplet hyperfine structure and $\tau$ is the optical depth. 

The fluctuation in $\dTb$ can be a powerful probe of the matter power spectrum when almost all the hydrogen atoms remain neutral after the recombination epoch and the spin temperature is much larger than the CMB temperature \citep[e.g.][]{McQuinn:2006}. 
It is possible that nature allows such a regime, called the ``X-ray heating epoch'', when the IGM is well heated above the CMB temperature by X-ray sources and the spin temperature of the hyperfine states is strongly coupled to the kinetic temperature of the baryonic gas through the efficient Lyman $\alpha$ scattering process (e.g. \citealt{Ahn:2015} and references therein). 
In addition, during this epoch, the cosmic reionization process of gas was not active enough to make a significant change in the neutral fraction of hydrogen atoms (but see, e.g., \citealt{Mirocha:2017} for a contrasting possibility that the X-ray heating becomes efficient only after the cosmic reionization process commences).
In this case, $\dTb$ is well approximated by\footnote{$\dTb$ depends on $\Omega_m$ and $\Omega_b$; we adopt the best-fit values for the Planck data.}
\begin{equation}
\dTb = 30.5 {\rm mK} \left(\frac{1+z}{10}\right)^{1/2}(1+\delta)(1-x),
\label{eq:dTb_approx}
\end{equation}
which indicates that when the ionized fraction $x$ is negligible $\dTb$ is proportional to the underlying density $\rho=\bar{\rho}(1+\delta)$.

For simplicity, we do not include the impact of the peculiar velocity on the observed $\dTb$ and $\pk$. 
While the observed power spectrum would be anisotropic or dependent on the direction of the wavenumber ${\bf k}$, it is a simple function of the line-of-sight component $\mu\equiv k_{}/{\bf k}$ and the isotropic 3D power spectra in the linear-density regime \citep[e.g.,][]{Barkana:2004zy,Mao:2012}. 
Therefore, our model dependence is solely reflected in the isotropic 3D power spectrum. We further assume that baryons follow the motion of the CDM such that $\delta$ is common to both of them, which is indeed a good approximation for large-scale modes in the matter-dominated era. 
Therefore, for the pre-reionization X-ray heating epoch, the 3D power spectrum of $\dTb$ is given by
\begin{equation}
P_{\dTb \dTb}(k)=(30.5 {\rm mK})^2 \left(\frac{1+z}{10} \right) P_{\delta \delta}  .
\label{eq:Pk_dtb}
\end{equation}
Of course, we note that there still remain many uncertainties due to the lack of any direct observations of the epoch of reionization. The X-ray background intensity is very uncertain \citep{Pritchard:2012}, and the degree of global ionized fraction, $\left< x \right>$, is very model-dependent \citep[e.g.][]{Iliev:2007,Ahn:2012,Wang:2015}.

\begin{figure}[hbt]
\epsscale{1.15}
\plottwo{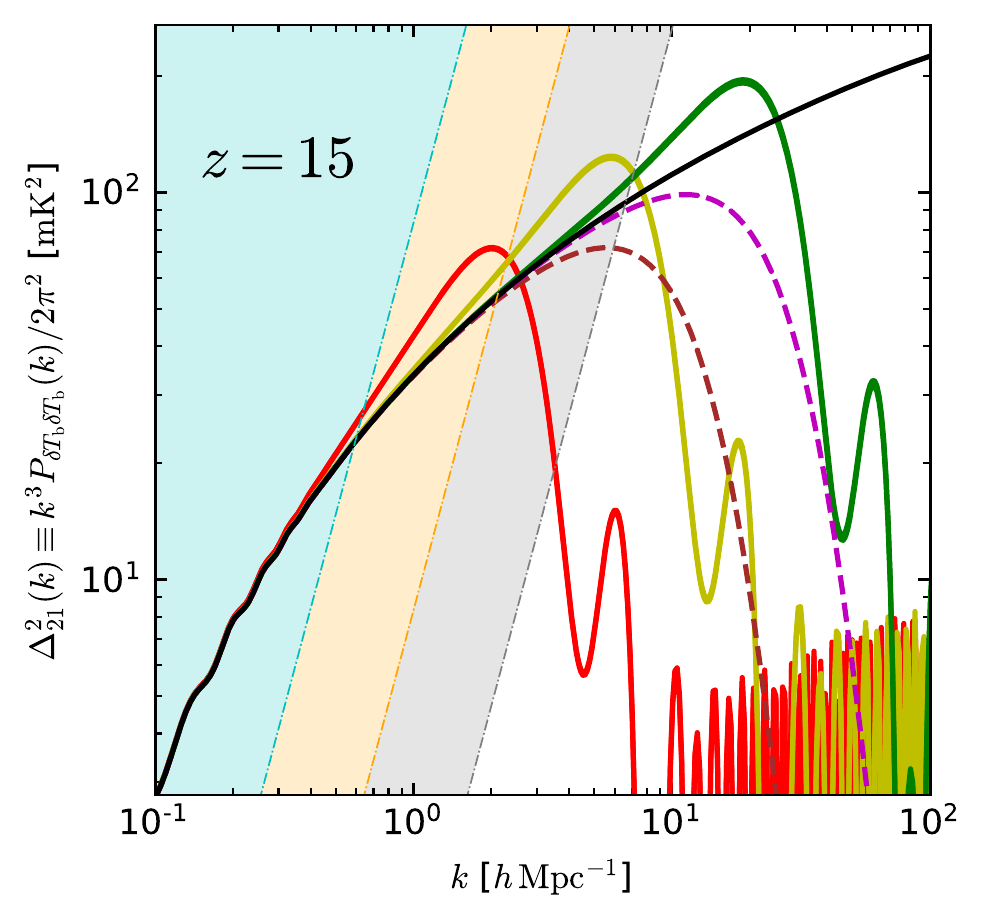}{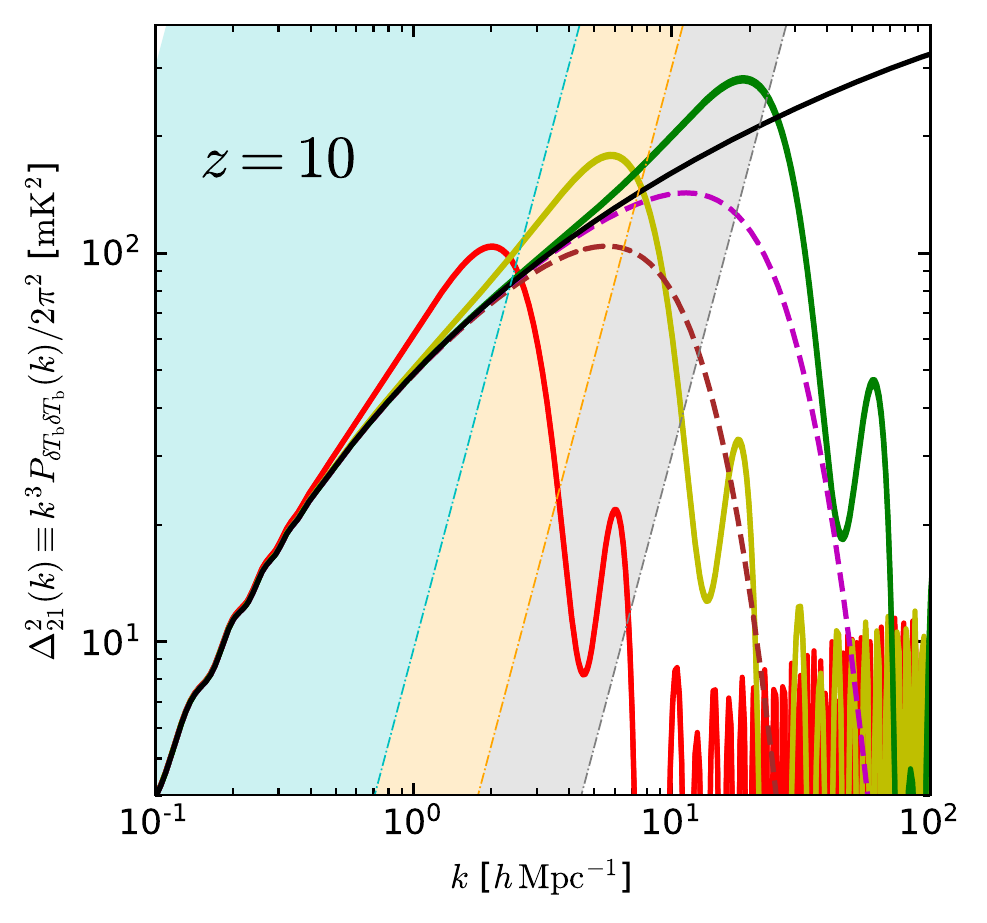}
\caption{The 21-cm power spectra $\Delta_{21}^2(k) \equiv k^3 P_{\dTb \dTb}(k)/2\pi^2$ of thermal inflation and WDM scenarios (colors) just before the epoch of reionization.
Shaded regions: the power spectra above the thermal noise from the modified SKA configuration with $100~{\rm deg^2}$ sky coverage.
Exposure times are ${\color{cyan}10^3}$, ${\color{orange}10^4}$, and ${\color{gray}10^5}$ hours in SKA1-LOW, and ${\color{cyan}10^2}$, ${\color{orange}10^3}$, and ${\color{gray}10^4}$ hours in SKA2-LOW \citep{Greig:2015}.
The upturn of the noise power spectrum at $k \sim 0.1 \invMpch$ \citep{Greig:2015} is not shown in this plot, because our main interest lies at $k \gtrsim 1 \invMpch$.}\label{21cmpower}
\end{figure}

\fref{21cmpower} shows the 21-cm power spectra $\Delta_{21}^2(k) \equiv k^3 P_{\dTb \dTb}(k)/2\pi^2$ from thermal inflation and WDM scenarios at $z = 15$ and $10$.
For understanding the observability of such power spectra, we also plot the power spectrum of the thermal noise, based on the SKA1-LOW configuration suggested by \citet{Greig:2015}. We assume a universal sky coverage of $100~{\rm deg^2}$  and $10^3$, $10^4$, and $10^5$-hour integrations, while the actual observation strategy may differ.
If reionization starts at relatively late epoch (e.g., $z \gtrsim 10$), then the 21-cm power spectrum for the thermal inflation scenario with $\kb \lesssim 3 \invMpc$ might be able to be distinguished from that for the standard $\Lambda$CDM scenario with $10^3$-hour exposure, by finding the enhancement of power spectra around $k \simeq \kb$ (see \fref{21cmpower}).
On the other hand, the WDM scenario with $\mFD \gtrsim 1 \keV$ cannot be distinguished from the standard $\Lambda$CDM scenario from $10^3$-hour exposure with SKA1-LOW.
If $10^4$-hour exposure is available, then one may be able to distinguish the thermal inflation scenario with $\kb \lesssim 10 \invMpc$ (and the WDM scenario with $\mFD \lesssim 1 \keV$) from the standard $\Lambda$CDM scenario, by capturing their characteristic features.

If reionization starts at earlier epoch (e.g., $z \gtrsim 15$), then $\Delta_{21}^2(k) $ at $z=10$ may be strongly contaminated by the reionization signal instead of reflecting the cosmological fluctuation. If so, one should instead observe the 21-cm background at $z\sim 15$. At that epoch,
both the thermal inflation scenario with $\kb \gtrsim 1 \invMpc$ and WDM scenario with $\mFD \gtrsim 1 \keV$ cannot be distinguished from the standard $\Lambda$CDM scenario from $10^3$-hour exposure of SKA1-LOW (see \fref{21cmpower}).
Even assuming $10^4$-hour exposure, which may be somewhat optimistic, only the thermal inflation scenario with $\kb \simeq 1 \invMpc$ can be distinguished from the standard $\Lambda$CDM scenario.  

We find that the 21-cm power spectrum of thermal inflation scenarios can be clearly distinguished from that of WDM scenarios, not to mention from that of the standard $\Lambda$CDM scenario, as long as $\kb$ is in the observable window. This is a merit we do not find in the global (see \sref{sec:massfun}) or local (see \sref{sec:milkyway}) halo mass functions.
This is mainly because the 21-cm observation is capable of observing the unprocessed power spectrum directly, as long as the right astrophysical condition is met. $\Delta_{21}^2(k) $ of thermal inflation scenarios shows a clear enhancement at $k \simeq \kb$, as well as the steep suppression at $k \gtrsim 3\kb$, which is quite distinct from the feature seen in WDM or $\Lambda$CDM scenarios.
This feature becomes more important for the detection of high-$\kb$ within the observation limit, e.g., for the detection of $\kb \simeq 3 \invMpc$ for $10^4$-hour exposure at $z =10$. 

The observational prospect becomes more optimistic if we consider the final telescope phase, SKA2-LOW, which would be available around the mid-2020s. SKA2-LOW is expected to have sensitivity about at least 10 times as much as that of SKA1-LOW \citep{Mellema:2015}.
If so, the observable boundary of $\kb$ in thermal inflation scenarios would increase to $\sim \times 3$ to that of SKA1-LOW with the same amount of integration. 
In the optimal case of SKA2-LOW, one then may be able to distinguish standard $\Lambda$CDM scenario from thermal inflation scenarios with $\kb \lesssim 10 \invMpc$ and $30 \invMpc$ by $10^3$ and $10^4$-hour integrations, respectively. In the presence of mixed models on the X-ray heating epoch, it is possible that an astrophysical source (e.g. cosmological H II regions during cosmic reionization) can contribute substantially to the observed 21-cm power spectrum. Even so, in a relatively early ionization state, it would be possible to separate the cosmological signal from the astrophysical one by the well-known $\mu$-decomposition scheme \citep{Barkana:2004zy,Mao:2012}, which seems optimistic especially with the strong feature seen in some thermal inflation scenarios.

\section{Summary/Discussion}\label{sec:summary}
In this paper, we studied how thermal inflation impacts the formation of cosmological structures and presented possible observational methods to probe the thermal inflation scenario. This scenario can be conveniently parametrized by a characteristic wavenumber $\kb$, which is determined by the number of e-foldings of the Universe during thermal inflation. 
The matter density fluctuation is boosted from that by the standard $\Lambda$CDM scenario at $k \simeq \kb$ and is strongly suppressed at  $k \gtrsim 3 \kb$ with rapid oscillation in $k$. 
Thus, observations should be focused on $k \gtrsim \kb$, and toward this end, we suggested three different observational targets: the global halo abundance at high redshifts, the abundance of galactic satellites of the Milky Way, and fluctuation in the 21-cm hydrogen radiation background before the epoch of reionization.
 
The 21-cm observation seems the most promising because the power spectrum of the 21-cm background fluctuation can become identical to the power spectrum of underlying matter density fluctuation. In the optimal case of low-ionization high-temperature of hydrogen atoms, and strong coupling between the spin temperature of the hyperfine states and the kinetic temperature of the baryonic gas, such a linear proportionality can exist. Depending on models of cosmic reionization, $z=10-15$ can be the target redshift for such a cosmological 21-cm observation. High-sensitivity radio telescope SKA will be able to probe thermal inflation scenarios with $\kb\sim 1 - 30 \invMpc$ progressively as its construction phase matures if $\sim$10000-hr integration is taken. 21-cm observation is expected to even distinguish between the thermal inflation scenario and the WDM scenario because they reflect their nature very differently in the power spectrum.

The other two observational methods may give some hints on the thermal inflation scenario because both the global and local halo abundances show deficits when compared to that of the standard $\Lambda$CDM scenario. 
However, these seem less promising than the 21-cm observation, because the observed objects are galaxies which must have gone through various nonlinear baryonic processes. In addition, the distinction between the thermal inflation scenario and the WDM scenario is not as distinct as in the 21-cm observation. Nevertheless, there can be other indirect consequences of such a deficit of small-mass halos, so these observations also need to be considered seriously. For example, the deficit of small-mass halos can delay cosmic reionization because of the induced deficit of radiation sources, and this may also reduce the gamma-ray background from dark matter annihilation which occurs at high-density regions such as the inner part of dark matter halos.

We note that, while not considered in the main body of the paper, the Lyman-$\alpha$ forest tomography can also constrain small-scale matter power spectrum.
For example, \citet{Baur:2016} suggests a lower bound of the mass of the warm dark matter from the observed high-resolution spectra of quasars at $z \gtrsim 4$ with the help from hydrodynamics simulations.
While further studies are required to fully understand the relation between the baryonic mass density along the line-of-sight and the amount of absorption in the quasar spectrum, the Lyman-$\alpha$ forest tomography by using the current or near-future high-resolution spectrographs may constrain the matter power spectrum up to $k \simeq 10 \invMpc$, which may constrain thermal inflation scenario up to $\kb \lesssim 10 \invMpc$.

There exist improvements to be made in our study. 
In studying the satellite galaxy abundance in the Milky Way, we used a combination of semi-numerical simulation code \textsc{pinocchio} and the one-to-one correspondence galaxy assignment formalism by \citet{Hong:2016}. This is a fast-track, approximate method compared to brute-force simulations using an $N$-body+hydrodynamics code, so our current prediction may be either confirmed or improved only after such expensive but more accurate calculations. One also needs to improve on the baryonic physics inside the Milky Way, because the photoheating and ionization of baryonic gas due to cosmic reionization can also reduce the observed number of Milky Way satellites. 

\acknowledgments
The authors thank Ewan Stewart and Donghui Jeong for helpful discussions. 
This work was supported by a research grant from Chosun University (2015). 

\software{\textsc{camb}\citep{Lewis:1999,Howlett:2012}, \textsc{pinocchio}\citep{Monaco:2002}}

\bibliographystyle{aasjournal}
\bibliography{21cm_and_thermal_inflation}

\end{document}